\documentclass[twocolumn]{aastex61}
\pdfoutput=1

\usepackage{soul}

\received{June 29, 2017}
\accepted{May 08, 2018}
\submitjournal{ApJ}

\shorttitle{Investigating Galactic SNR candidates with LOFAR}
\shortauthors{Driessen et al.}

\begin{document}

\title{Investigating Galactic supernova remnant candidates using LOFAR}

\correspondingauthor{Laura N. Driessen}
\email{Laura@Driessen.net.au}

\author{Laura N. Driessen}
\affiliation{Anton Pannekoek Institute for Astronomy, University of Amsterdam, Science Park 904, 1098 XH Amsterdam, The Netherlands}
\affiliation{Jodrell Bank Centre for Astrophysics, School of Physics and Astronomy, The University of Manchester, Manchester,  M13 9PL, UK}
\author{Vladim{\'i}r Dom{\v c}ek}
\affiliation{Anton Pannekoek Institute for Astronomy, University of Amsterdam, Science Park 904, 1098 XH Amsterdam, The Netherlands}
\affiliation{GRAPPA, GRavitation and AstroParticle Physics Amsterdam, University of Amsterdam, Science Park 904, 1098 XH Amsterdam, Netherlands}
\author{Jacco Vink}
\affiliation{Anton Pannekoek Institute for Astronomy, University of Amsterdam, Science Park 904, 1098 XH Amsterdam, The Netherlands}
\affiliation{GRAPPA, GRavitation and AstroParticle Physics Amsterdam, University of Amsterdam, Science Park 904, 1098 XH Amsterdam, Netherlands}
\affiliation{SRON, Netherlands Institute for Space Research, Sorbonnelaan 2, 3584 CA, Utrecht, The Netherlands}
\author{Jason W. T. Hessels}
\affiliation{Anton Pannekoek Institute for Astronomy, University of Amsterdam, Science Park 904, 1098 XH Amsterdam, The Netherlands}
\affiliation{ASTRON, Netherlands Institute for Radio Astronomy, Postbus 2, 7990 AA, Dwingeloo, The Netherlands}
\author{Maria Arias}
\affiliation{Anton Pannekoek Institute for Astronomy, University of Amsterdam, Science Park 904, 1098 XH Amsterdam, The Netherlands}
\author{Joseph D. Gelfand}
\affiliation{NYU Abu Dhabi, PO Box 129188, Abu Dhabi, UAE}
\affiliation{Affiliate Member, Center for Cosmology and Particle Physics, New York University, 726 Broadway, New York, NY 10003}

\begin{abstract}
{We investigate six supernova remnant (SNR) candidates -- $\mathrm{G}51.21+0.11$, $\mathrm{G}52.37-0.70$, $\mathrm{G}53.07+0.49$, $\mathrm{G}53.41+0.03$, $\mathrm{G}53.84-0.75$, and the possible shell around $\mathrm{G}54.1+0.3$ -- in the Galactic Plane using newly acquired LOw-Frequency ARray (LOFAR) High-Band Antenna (HBA) observations, as well as archival Westerbork Synthesis Radio Telescope (WSRT) and Very Large Array Galactic Plane Survey (VGPS) mosaics}. {We find that $\mathrm{G}52.37-0.70$, $\mathrm{G}53.84-0.75$, and the possible shell around pulsar wind nebula $\mathrm{G}54.1+0.3$ are unlikely to be SNRs, while $\mathrm{G}53.07+0.49$ remains a candidate SNR. $\mathrm{G}51.21+0.11$ has a spectral index of $\alpha=-0.7\pm0.21$, but lacks X-ray observations and as such requires further investigation to confirm its nature.}  We confirm one candidate, $\mathrm{G}53.41+0.03$, as a new SNR because it has {a shell-like morphology,} a radio spectral index of $\alpha=-0.6\pm0.2$ and it has the X-ray spectral characteristics of a $1000-8000$ year old SNR. The X-ray analysis was performed using archival {\it XMM-Newton} observations, which show that $\mathrm{G}53.41+0.03$ has strong emission lines and is best characterized by a non-equilibrium ionization model, consistent with an SNR interpretation.  Deep Arecibo radio telescope searches for a pulsar associated with $\mathrm{G}53.41+0.03$ resulted in no detection, but place stringent upper limits on the flux density of such a source if it is beamed towards Earth.
\end{abstract}


\section{Introduction}

There are many shell- and bubble-like objects in our Galaxy. {For example, there are 295 supernova remnants (SNRs) in Green's SNR catalog \citep[just under half of which have only been detected in the radio;][]{2014BASI...42...47G,Green2017}, 76 SNR candidates in a recent THOR+VGPS analysis \citep{2017arXiv170510927A}, and $\sim1500$ known HII regions (as well as $\sim2500$ probable and $\sim4000$ candidate HII regions) in the WISE HII region catalog\footnote{\href{http://astro.phys.wvu.edu/wise/}{www.astro.phys.wvu.edu/wise}} \citep{2014ApJS..212....1A}}. This means that observations and surveys of the Galactic Plane capable of investigating shell-like objects, particularly observations differentiating between candidate HII regions and SNRs, are extremely useful. {As there are many sources and candidates, targeting individual objects with a single pointing per object is impractical. Interferometers that can observe large areas of the sky at low-frequencies with wide frequency bandwidth should prove to be excellent tools for Galactic Plane investigations.}

About 90\% of SNRs and SNR candidates have been found in radio surveys \citep{2017arXiv170510927A}, but it is thought that there could be many missing SNRs \citep[e.g.][]{1991ApJ...378...93L,1994ApJS...92..487T,2014A&A...566A..76G}. Obtaining a more  complete record of the SNR population, including confirming or rejecting the nature of SNR candidates, is important as it leads to better estimates of the Galactic supernova rate, the maximum ages of SNRs, and because SNRs are obvious locations for searching for young pulsars.

Low-frequency ($\lesssim350\,\mathrm{MHz}$) Galactic Plane observations are useful for investigating SNRs and SNR candidates, particularly for differentiating between SNR candidates and HII regions, due to the typically steeper radio spectral indices of SNRs \citep[$\alpha\approx-0.5$;][]{2013Ap&SS.346....3O} as compared to HII regions ($\alpha\gtrsim 0$); where $S_{\nu}\,\propto\,\nu^{\alpha}$ for $S_{\nu}$ integrated flux density in $\mathrm{Jy}$ and $\nu$ frequency in $\mathrm{Hz}$ \citep{2013Ap&SS.346....3O}. This means that SNRs are brighter at lower frequencies, while HII regions are brighter at higher frequencies. {However, there have been relatively few low-frequency surveys of the Galactic Plane with high angular resolution and sensitivity.} A good illustration of the capability of such surveys for SNR searches was demonstrated by a $333\,\mathrm{MHz}$ survey with the Very Large Array (VLA) of the Galactic Center region \citep{2006ApJ...639L..25B}. This survey resulted in the discovery of 35 new candidate SNRs, {31 of which are now confirmed}. Multi-wavelength analysis is required to confirm (or reject) SNR candidates, such as X-ray observations or further radio observations at a different frequency to confirm the spectral index.

The LOw-Frequency ARray \citep[LOFAR;][]{2013A&A...556A...2V} is an interferometer that observes at low-frequencies with a large field-of-view 
{(FoV; e.g. $\sim 11\,\mathrm{deg^{2}}$ using HBA Dual Inner mode)},
which means that it is ideal for observing and discovering steep-spectra objects and for differentiating between SNR candidates and HII regions. LOFAR consists of two arrays: the Low Band Antennas (LBA) and the High Band Antennas (HBA). The LBA observes between $10$ and $80\,\mathrm{MHz}$ while the HBA observes between $110$ and $250\,\mathrm{MHz}$. The wide FoV also introduces many technical difficulties regarding calibration and imaging, particularly as the ionosphere can introduce significant phase and amplitude variations across the FoV.
 
Here we discuss six SNR candidates in the FoV of proprietary LOFAR HBA observations (PI: J.\,D.\,Gelfand) that overlap with an archival Westerbork Synthesis Radio Telescope (WSRT) mosaic \citep{1996ApJS..107..239T} and an archival VLA Galactic Plane Survey (VGPS) mosaic \citep{2006AJ....132.1158S}.
These SNR candidates were identified in a study of THOR+VGPS observations by \citet{2017arXiv170510927A} and are: $\mathrm{G}51.21+0.11$, $\mathrm{G}52.37-0.70$, $\mathrm{G}53.07+0.49$, $\mathrm{G}53.41+0.03$, $\mathrm{G}53.84-0.75$, and $\mathrm{G}54.1+0.3$. In particular, we present a multi-frequency analysis of SNR candidate $\mathrm{G}53.41+0.03$. In Section\,\ref{sec:Observations} we present the observations. {In Section\,\ref{sec: results} we present our results and in Section\,\ref{sec: discussion} we discuss the SNR candidates.}  We conclude in Section\,\ref{sec: conclusion}.

\section{Observations and Analysis}
\label{sec:Observations}

\subsection{Radio observations}
\label{sec: radio obs}

\begin{figure*}
\plotone{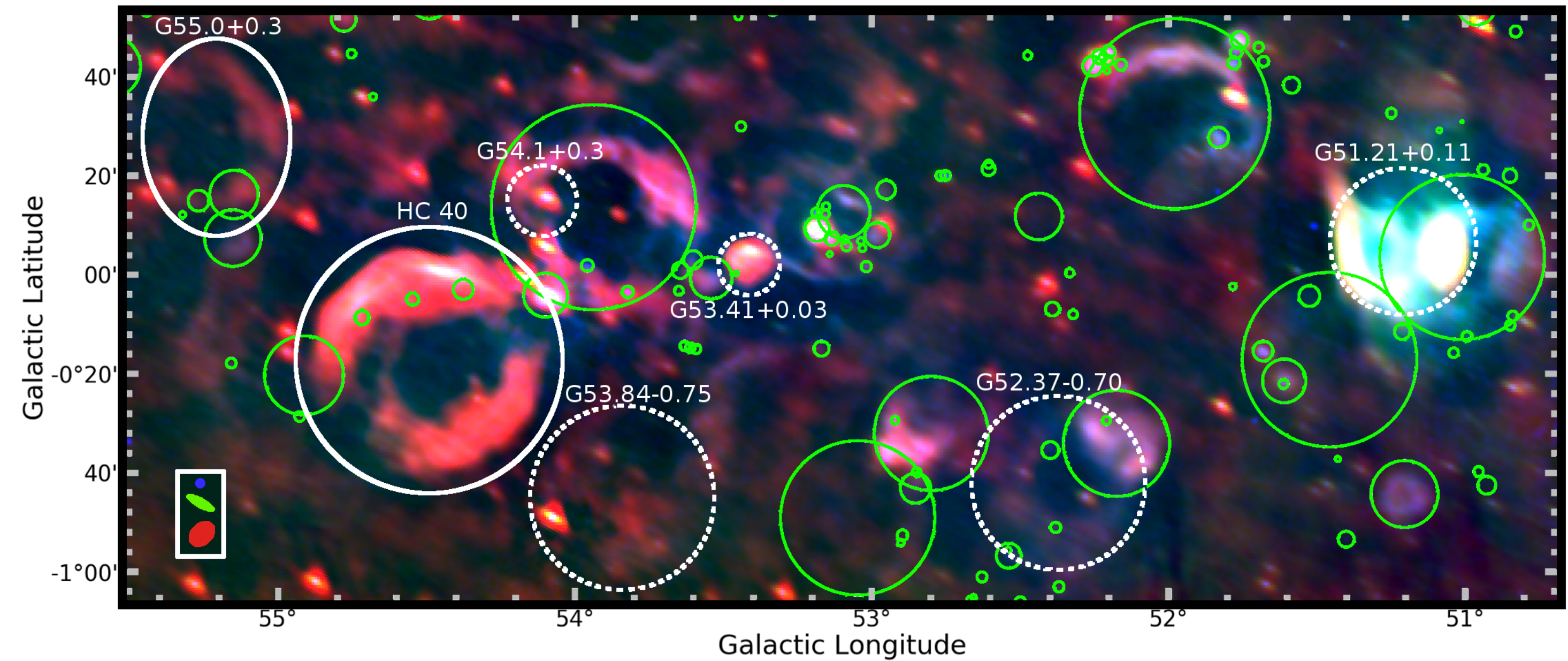}
\caption{Observations of the Galactic Plane at $1.4\,\mathrm{GHz}$ (blue, VLA), $327\,\mathrm{MHz}$ (green, WSRT), and $144\,\mathrm{MHz}$ (red, LOFAR HBA). The synthesized beam sizes are shown in the bottom left corner. Known SNRs from Green's SNR catalog \citep{2014BASI...42...47G} are circled in solid white and candidate SNRs from \citet{2017arXiv170510927A} are circled in dashed white. {The green circles are HII regions from the WISE catalog \citep{2014ApJS..212....1A}.}
} 
\label{fig: F1}
\end{figure*}

We use radio observations at three different frequencies -- $144\,\mathrm{MHz}$ (LOFAR HBA), $327\,\mathrm{MHz}$ (WSRT), and $1400\,\mathrm{MHz}$ (VGPS) -- to investigate part of the Galactic Plane.
Figure\,\ref{fig: F1} shows the FoV where our LOFAR HBA observations overlap archival WSRT and VGPS mosaics.

We initially obtained and analyzed the LOFAR observations to investigate pulsar wind nebula (PWN) $\mathrm{G}54.1+0.3$ but, due to the large FoV, we also investigated other promising SNR candidates.
The LOFAR observations are centered on the PWN. The observations were taken on 2015 June 12 as part of project LC4\_011 (ObsID: 345918) and were performed in HBA Dual Inner mode \citep{2013A&A...556A...2V}. This means that the inner 24 tiles of the remote stations were  used resulting in a full-width half-maximum (FWHM) of the primary beam of $3.8^{\circ}$ and FoV of $\sim11\,\mathrm{deg^{2}}$ {in this configuration}. The LOFAR HBA target and calibrator scans cover the frequency range from $118.7\,\mathrm{MHz}$ to $169.5\,\mathrm{MHz}$. The observing bandwidth was split into 260 subbands (SBs) with bandwidth of $195.3\,\mathrm{kHz}$ each. For these observations an 18 min calibrator scan of 3C380 was taken before and after the 3 hr target scan.

The LOFAR observations were flagged, demixed, and averaged as part of standard LOFAR pre-processing. Demixing involves removing the effects of the very bright radio sources, Cassiopeia A and Cygnus A, that affect LOFAR images even when they are far from the phase center of the FoV. The data were averaged to 4 frequency channels per SB. {The LOFAR synthesized beam size is $3.0'\times2.2'$ with a position angle of $220.7\deg$ (with respect to the Galactic Plane) at $144\,\mathrm{MHz}$ using a Briggs weighting of 1.0 \citep{1995AAS...18711202B}.} As this is a Galactic Plane observation the imaging calibration pipeline, prefactor \citep[or Pre-Facet-Cal;][]{2016ApJS..223....2V}, was not successful. This is due to the significant extended emission in the Galactic Plane, across the FoV. Ionospheric variations during the observations were particularly pronounced. The observation was calibrated by transferring the time-independent, zero-phase gain solutions from the second calibrator scan to the target scan. The observations were then summed into 26 measurement sets of 10 SBs each. Two rounds of self-calibration were then performed on the target scan, the first using a model from the TIFR GMRT Sky Survey (TGSS) Alternative Data Release \citep[TGSS ADR\footnote{\url{http://tgssadr.strw.leidenuniv.nl/doku.php}},][]{2017A&A...598A..78I}. Multiscale imaging with Briggs 1.0 weighting was then performed using the WSClean tool \citep{offringa-wsclean-2014}. The subband with a central frequency of $150\,\mathrm{MHz}$ was flux calibrated using the integrated flux density measurements of point sources from the TGSS ADR. It is important to note that the sensitivity of the LOFAR image drops significantly at the edge of the FWHM of the primary beam. This means that flux density values far from the phase center (PWN $\mathrm{G}54.1+0.3$) are less reliable. Figure \ref{fig: F1} was produced by performing a multi-frequency (MFS) clean on all measurement sets.

WSRT observations were obtained from a Galactic Plane point source survey at $327\,\mathrm{MHz}$ with a beam size of $60''\times191''$ {and a position angle of $61.3^{\circ}$ (with respect to the Galactic Plane)} by \citet{1996ApJS..107..239T}\footnote{\url{www.ras.ucalgary.ca/wsrt_survey.html}}.
VLA observations from VGPS with a beam size of $1'\times1'$ were also used \citep{2006AJ....132.1158S}\footnote{\url{www.ras.ucalgary.ca/VGPS/VGPS_data.html}}. The VLA observations have the highest angular resolution of the available radio observations of this FoV.

\subsection{Radio pulse search observations}
\label{sec: obs pulse search}

To search for a pulsar towards $\mathrm{G}53.41+0.03$, we observed the
region using the 305-m Arecibo radio telescope and the 7-beam Arecibo
L-band Feed Array (ALFA) receiver.  On 2017 June 21, we made a
3-pointing grid of the region, where together the 21 observed beams
were interleaved and cover a roughly $10^{\prime}$ region around the
center of $\mathrm{G}53.41+0.03$.  The first pointing, where the central beam of ALFA
was directly pointed {towards the apparent center of} $\mathrm{G}53.41+0.03$,
integrated for 2400\,s.  The other two
interleaving pointings were integrated for 900\,s.  We recorded the
resulting filterbank data using the Mock spectrometers, which provided
two partially overlapping 172\,MHz subbands centered at 1300 and
1450\,MHz, respectively.  Only total intensity was recorded, with 0.34-MHz spectral
channels and 65.5\,$\mu$s time resolution.  We converted the raw
samples from 16-bit to 4-bit values subsequent to the observation in
order to reduce the data volume.  At the start of the session, we
observed PSR~J1928+1746 in the central ALFA beam, in order to verify the
configuration.

We searched for radio pulsations in the direction of $\mathrm{G}53.41+0.03$ using standard methods, as
implemented in the PRESTO\footnote{\url{https://github.com/scottransom/presto}} software package.  We chose
to search the Mock subbands separately because the lower-frequency
subband contains significantly more radio frequency interference
(RFI).  For each beam and subband we excised RFI using {\tt rfifind}
and then used multiple calls to {\tt prepsubband} to generate
dedispersed time series for dispersion measures in the range DM = $0 - 1019\,\mathrm{pc}\,\mathrm{cm}^{-3}$ in steps of 
$1\,\mathrm{pc}\,\mathrm{cm}^{-3}$.  The remaining dispersive smearing is $\sim 1$\,ms, even for the highest DMs in this range.  Each dedispersed
timeseries was then searched for periodicities using {\tt accelsearch}
with no additional search for linear acceleration (i.e. $z_{\rm max} = 0$).  The cumulative set of candidates was
then sifted and ranked using {\tt ACCEL\_sift.py}.  We folded promising
candidates --- those with high signal-to-noise, high coherent power,
and apparent peaks in signal-to-noise as a function of DM --- using
{\tt prepfold}.  Associated diagnostic plots for each candidate were
then visually inspected.  When {this approach was} applied to the test pulsar, J1928+1746, the
expected signal was easily recovered in both subbands.

\subsection{Infrared observations}
The FoV coinciding with the radio observations was observed at $24.0\,\mathrm{\mu m}$ as part of the Multiband Infrared Photometer for Spitzer GALactic Plane (MIPSGAL) survey \citep{2009PASP..121...76C,2015AJ....149...64G}. MIPSGAL $24.0\,\mathrm{\mu m}$ observations have a resolution of $6''$ and a $5\sigma$ root-mean-squared sensitivity of $1.3\,\mathrm{mJy}$. 

\subsection{X-ray observations}
\label{xray_obs}

Of the six candidate SNRs that we investigate in this paper, only the possible shell around PWN\,$\mathrm{G}54.1+0.3$ has been analysed previously in the X-ray band. It has been observed using {\it Chandra} \citep{2002ApJ...568L..49L}, {\it Suzaku}, and {\it XMM-Newton} \citep{2010A&A...520A..71B}.

The position of $\mathrm{G}53.84-0.75$ has been observed in a {\it ROSAT} PSPC observation (ObsID: WG500209P.N1). Using the region size of $18.7\,'$ \citep{2017arXiv170510927A} we estimate the X-ray count rate with $2\sigma$ upper limit to be $1.5\times 10^{-2}$ counts/sec in the {\it ROSAT} $0.4$ -- $2.4\,\mathrm{keV}$ energy band. 

$\mathrm{G}53.41+0.03$ is detected at the edge of the FoV of two {\it ROSAT} PSPC observations (ObsIDs: WG500042P.N2 and WG500209P.N1) and partially covered by an \textit{XMM-Newton} observation taken on  2008 Mar 29 (ObsID: 0503740101).  The other three SNR candidates, $\mathrm{G}51.21+0.11$, $\mathrm{G}52.37-0.70$, and $\mathrm{G}53.07+0.49$, have no complementary data available in the X-ray band.

\begin{figure}
\plotone{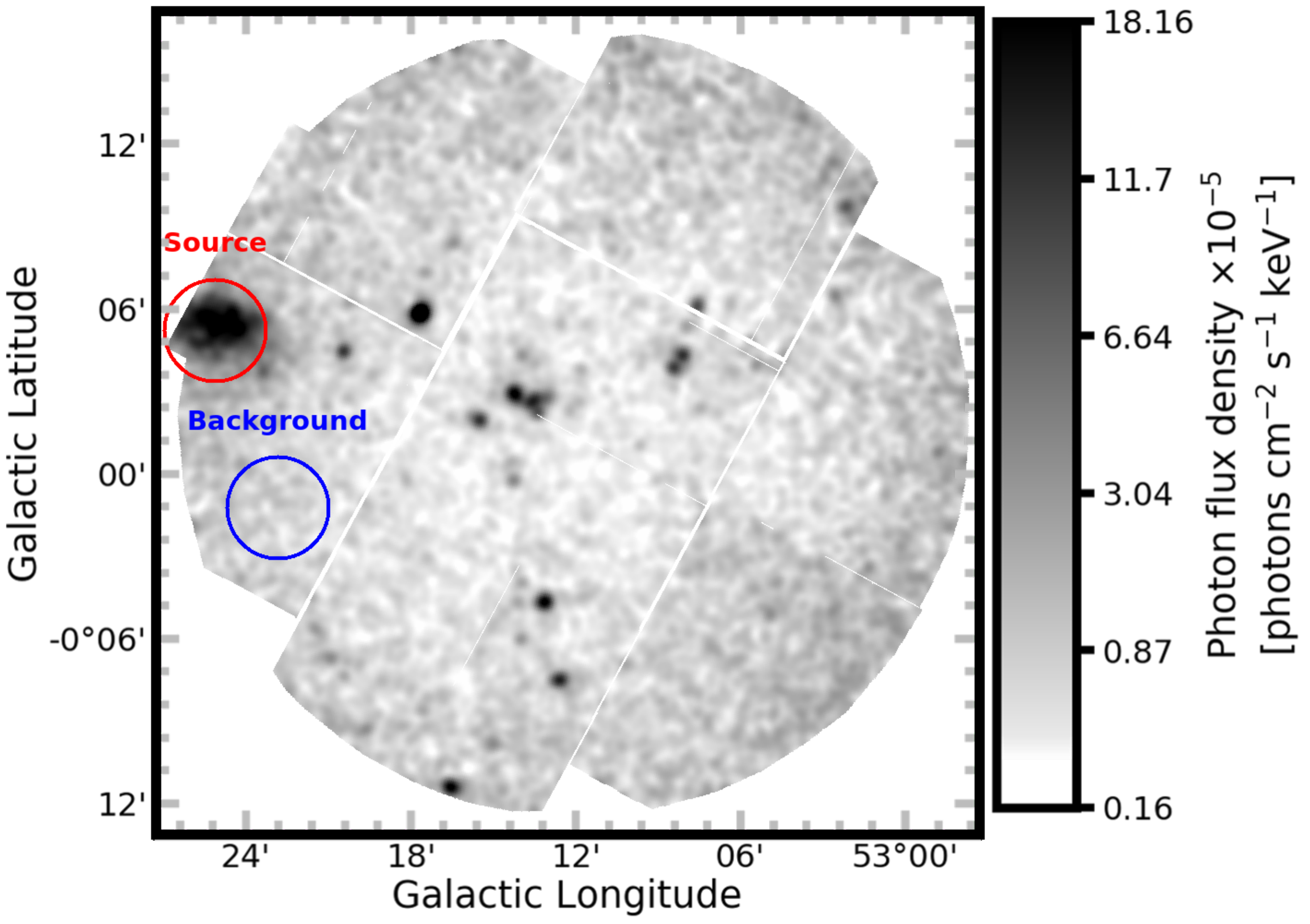}
\caption{{Exposure and vignetting corrected image from the {\it XMM-Newton} MOS2 detector.} Source and background extraction regions are circled.
} 
\label{fig: F2}
\end{figure}

Although $\mathrm{G}53.41+0.03$ lies at the edge of the
detector in the \textit{XMM-Newton} observation,
the observation is important as it allows us to determine the nature of the X-ray emission through spectral analysis of the EPIC-MOS camera \citep{Turner2000} data. We extracted the  spectrum with the Science Analysis System (SAS) v14.0. Due to a failed CCD chip in MOS1 and the smaller FoV of the EPIC-PN detector only data from the MOS2 detector were used. The data were reduced using the \texttt{emproc} task and filtered for the background flaring. This resulted in $40.7\,\mathrm{ks}$ of cleaned exposure time. The source extraction region was a $1.8\,'$ radius circle centered on the extended X-ray source. The background was extracted using a region of the same size positioned in a nearby area of the detector devoid of X-ray sources.  The source and background regions are shown in Figure \ref{fig: F2}. 

To perform the spectral analysis the SPEX fitting package version 3.04 (2017) together with SPEXACT 2.07 atomic tables were used \citep{Kaastra1996}. The fitting statistics method employed was C-statistics \citep{Cash1979}. Abundances were expressed with respect to Solar Abundance values
of \citet{Lodders2009}. For the emission measure parameter ($n_\mathrm{e}n_\mathrm{H}V$) we assumed a distance of $7.5\,\mathrm{kpc}$ (see Sec.\,\ref{sec: G53 distance}). The analysis of the spectra was performed in the energy range between $0.7$ -- $3.0\,\mathrm{keV}$, as this is the range in which the source spectrum dominates the background. The \texttt{Obin} command was used to obtain optimal binning of the spectra. After background subtraction the source spectrum consists of $\sim 2000$ counts. 
The spectrum was fit with a non-equilibrium ionization (NEI) model with Galactic absorption. The Galactic background was represented by the model \texttt{hot} in SPEX, with the temperature fixed to $0.5\,\mathrm{eV}$ to mimic absorption by neutral gas \citep{SpexCookbook}. The NEI model was employed with the following free parameters: electron temperature $T_2$, ionization age $ \tau = n_e t $, normalization $n_e n_H V$, and abundances of elements Ne, Mg, Si, S, Fe. These elements have line emission in the energy band from $0.8$ -- $2.6\mathrm{keV}$, the band for which there was sufficient signal to noise. 

\subsection{High-energy observations}
\label{sec: high energy obs}
We searched the High Energy Stereoscopic System CATalog (HESSCAT\footnote{\href{https://www.mpi-hd.mpg.de/hfm/HESS/pages/home/sources/}{www.mpi-hd.mpg.de/hfm/HESS/pages/home/sources/}}) and Third {\it Fermi} LAT Catalog of High-Energy Sources \citep[3FGL;][]{2015ApJS..218...23A} for high-energy sources associated with any of the SNR candidate shells. {\it Fermi} source 3FGL J1931.1+1659 is within the radius of SNR candidate $\mathrm{G}52.37-0.70$.
There are no other high-energy sources close to the other five SNR candidates.

\section{Results}
\label{sec: results}

\begin{figure*}
\plotone{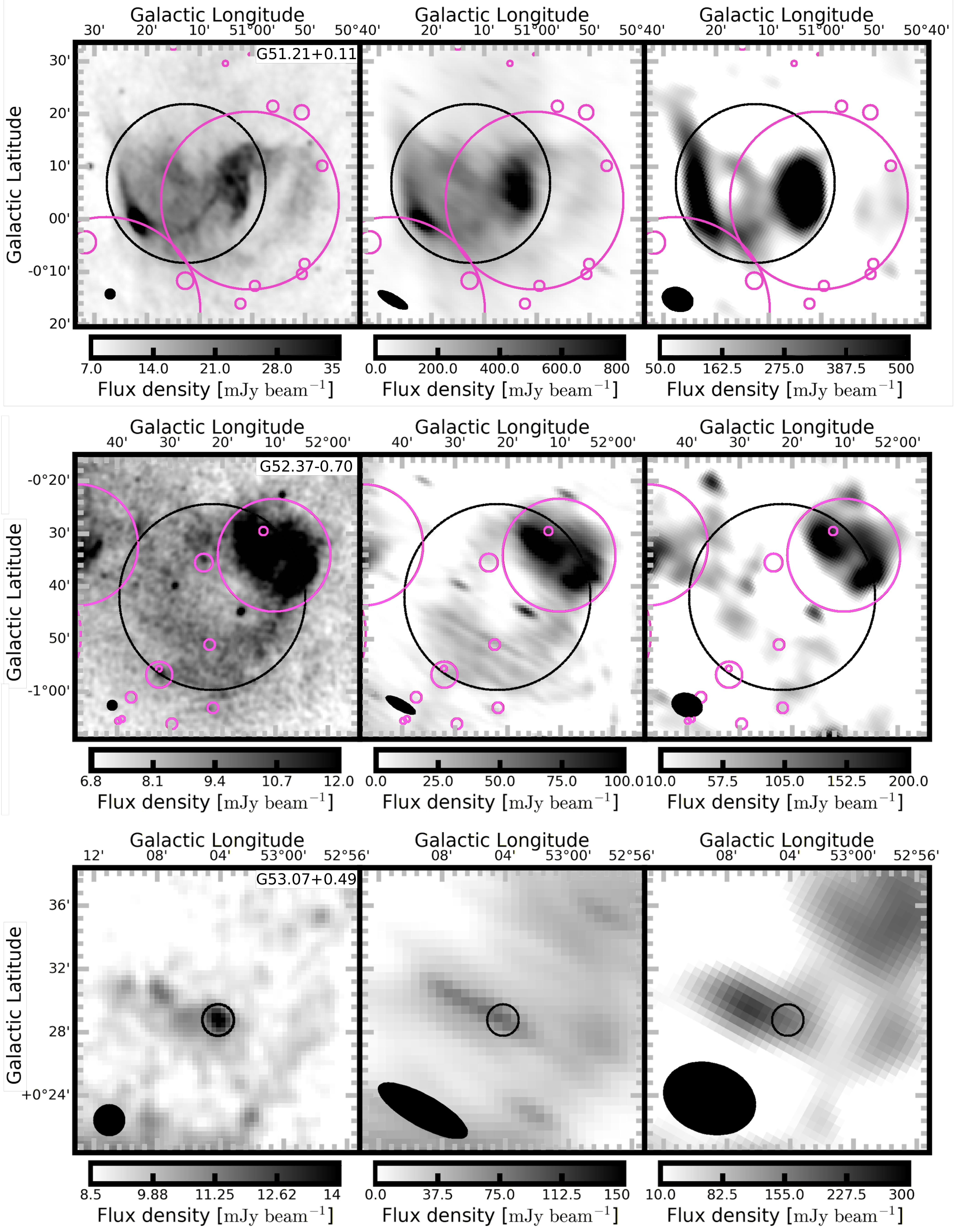}
\caption{SNR candidates from \citet{2017arXiv170510927A} in the FoV. In each row the left panel is the VGPS $1.4\,\mathrm{GHz}$ observation, the center panel is the WSRT $327\,\mathrm{MHz}$ observation, and the right panel is the LOFAR $144\,\mathrm{MHz}$ observation. From top to bottom the rows are the SNR candidates (circled in black) from \citet{2017arXiv170510927A}: $\mathrm{G}51.21+0.11$, $\mathrm{G}52.37-0.70$, and $\mathrm{G}53.07+0.49$. {The magenta circles are HII regions from the WISE HII catalog \citep{2014ApJS..212....1A}.} The beam sizes are shown in the bottom left corner of each panel. \label{fig: F3}}
\end{figure*}

\begin{figure*}
\plotone{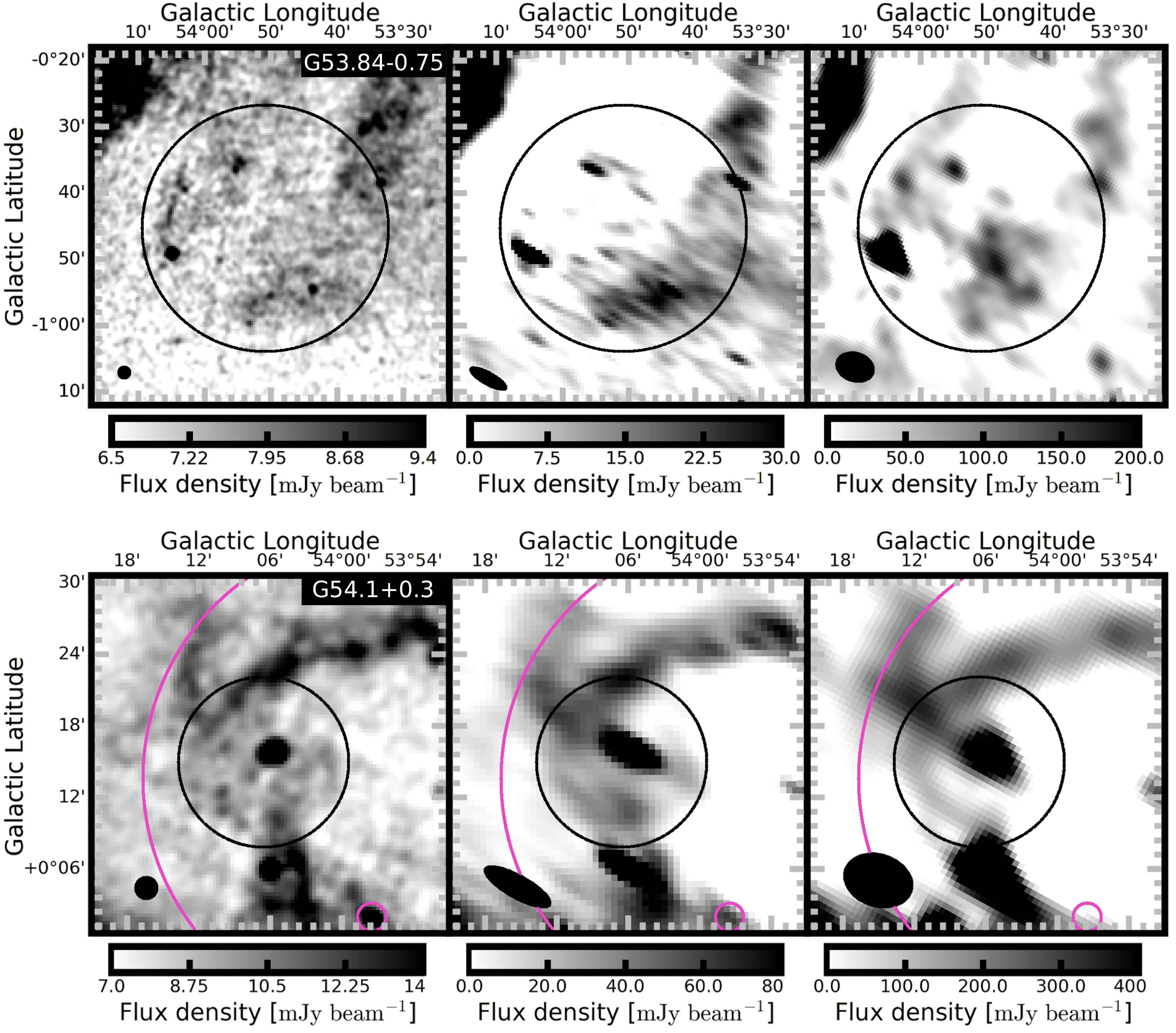}
\caption{SNR candidates from \citet{2017arXiv170510927A} in the FoV. In each row the left panel is the VGPS $1.4\,\mathrm{GHz}$ observation, the center panel is the WSRT $327\,\mathrm{MHz}$ observation, and the right panel is the LOFAR $144\,\mathrm{MHz}$ observation. From top to bottom the rows are the SNR candidates (circled in black) from \citet{2017arXiv170510927A}: $\mathrm{G}53.84-0.75$ and $\mathrm{G}54.1+0.3$. {The magenta circles are HII regions from the WISE HII catalog \citep{2014ApJS..212....1A}.} The beam sizes are shown in the bottom left corner of each panel. \label{fig: F4}}
\end{figure*}

The VGPS, WSRT, and LOFAR HBA observations of the six SNR candidates in the FoV -- $\mathrm{G}51.21+0.11$, $\mathrm{G}52.37-0.70$, $\mathrm{G}53.07+0.49$, $\mathrm{G}53.41+0.03$, $\mathrm{G}53.84-0.75$, and $\mathrm{G}54.1+0.3$ -- are shown in Figures \ref{fig: F3} and \ref{fig: F4}. Only $\mathrm{G}53.41+0.03$ and $\mathrm{G}54.1+0.3$ have been observed in the X-ray band (see Sec.\,\ref{xray_obs}). {As discussed by \citet{2017arXiv170510927A}, all six of the candidates have low thermal emission compared to the non-thermal emission, which we confirm using the MIPSGAL observations.}

\subsection{Radio results}

The flux densities and spectral indices of SNR candidates $\mathrm{G}51.21+0.11$, $\mathrm{G}52.37-0.70$, $\mathrm{G}53.41+0.03$, and $\mathrm{G}53.84-0.75$, and the candidate shell around PWN $\mathrm{G}54.1+0.3$ measured using the positions and radii reported by \citet{2017arXiv170510927A} are shown in Table\,\ref{tab: T1}. We subtracted the integrated flux density of the HII region overlapping $\mathrm{G}52.37-0.70$ and the flux density of the bright point source within $\mathrm{G}53.84-0.75$.  Due to a drop-off in sensitivity away from the phase center of the HBA observation, we do not measure LOFAR integrated flux densities for $\mathrm{G}51.21+0.11$ and $\mathrm{G}52.37-0.70$.

\begin{table*}
\centering
\begin{tabular}{l|lll|l}
 & \multicolumn{3}{c|}{Flux density (Jy)} & \\
\cline{2-4}
SNR & $1.4\,\mathrm{GHz}$ & $327\,\mathrm{MHz}$ & $150\,\mathrm{MHz}$ & $\alpha$ \\
\hline
\hline
\vspace{-5mm}
\\
G51.21+0.11 & $24.35\pm2.1$ & $66.1\pm0.1$ &  & $-0.7\pm0.21$ \\
G52.37$-$0.70 & $5.24\pm1.75$ & $3.2\pm0.03$ &  & $0.3\pm0.3$ \\
G53.41+0.03 & $1.21\pm0.21$ & $2.2\pm0.03$ & $3.11\pm0.2$ & $-0.6\pm0.2$ \\
G53.84$-$0.75 & $1.31\pm3.43$ & $0.06\pm0.02$ & $1.2\pm0.07$ & $0.05\pm3.9$ \\
G54.1+0.3 & $1.46\pm0.28$ & $1.21\pm0.05$ & $0.4\pm0.8$ & $0.3\pm4.3$\\
\hline
\hline
\end{tabular}
\caption{Integrated flux densities of SNR candidates at $1.4\,\mathrm{GHz}$ from \citet{2017arXiv170510927A}, $327\,\mathrm{MHz}$ measured using WSRT observations \citep{1996ApJS..107..239T}, and $150\,\mathrm{MHz}$ using LOFAR HBA observations. The WSRT errors are $3\sigma$ statistical errors based on the RMS noise in the image; these errors do not take other sources of error, such as confusion, into account. $\alpha$ was obtained {using a simple power law and} a weighted least squares fit using the measured error plus 20\% for systematics. {We note that a simple power law is not always the best model, for example for $\mathrm{G}54.1+0.3$.}}
\label{tab: T1}
\end{table*}

{SNR candidate $\mathrm{G}51.21+0.11$, shown in Figure.\,\ref{fig: F3} (top row), has a complex morphology with a bright radio filament type structure and a bright radio patch. It has an HII region, $\mathrm{G}051.010+00.060$ \citep{2014ApJS..212....1A}, on one side that appears to be coincident.}

{$\mathrm{52.37-0.70}$ is a faint radio shell visible most clearly in the VLA observation in the second row of Figure\,\ref{fig: F3}. There is a bright HII region, $\mathrm{G}052.174-00.567$ \citep{2014ApJS..212....1A}, on the upper right of this candidate and some smaller HII regions within the shell.}

$\mathrm{G}53.07+0.49$ has a small angular size \citep[a radius of only $1'$, ][]{2017arXiv170510927A} and the location of the peak flux density is different for WSRT and LOFAR compared to the original VLA identification of the candidate. In Figure \ref{fig: F3} (bottom panel) we can also see that there is some extended emission around $\mathrm{G}53.07+0.49$ that may or may not be associated with this candidate. {As it is unclear which emission in the WSRT and LOFAR observations may or may not be associated with the candidate we do not measure WSRT or LOFAR flux densities for this candidate.}

{There is diffuse emission and some radio point sources in the region where candidate $\mathrm{G}53.84-0.75$ is located (Fig.\,\ref{fig: F4},\,upper panel), but it is difficult to identify what emission is related to candidate $\mathrm{G}53.84-0.75$ and whether there is a discrete object or if the extended emission is Galactic Plane dust.}

{PWN $\mathrm{G}54.1+0.3$ is shown in Figure\,\ref{fig: F4}\,(lower panel) where the bright spot in the center is the PWN and the partial loop around it is the known HII region $\mathrm{G}053.935+00.228$ \citep{2014ApJS..212....1A}. There is some faint, diffuse radio emission around the PWN in the VLA observation, which is the SNR-shell candidate.
In the WSRT and LOFAR observations of PWN $\mathrm{G}54.1+0.3$ shown in Figure\,\ref{fig: F4} it appears that the possible shell identified in the VLA observations \citep{2017arXiv170510927A} fades away or is part of the surrounding HII region. The large uncertainty in the spectral index in Table\,\ref{tab: T1} reflects that a powerlaw is not the best model; however, the flux density clearly decreases as the frequency decreases.}

\begin{figure*}
\plotone{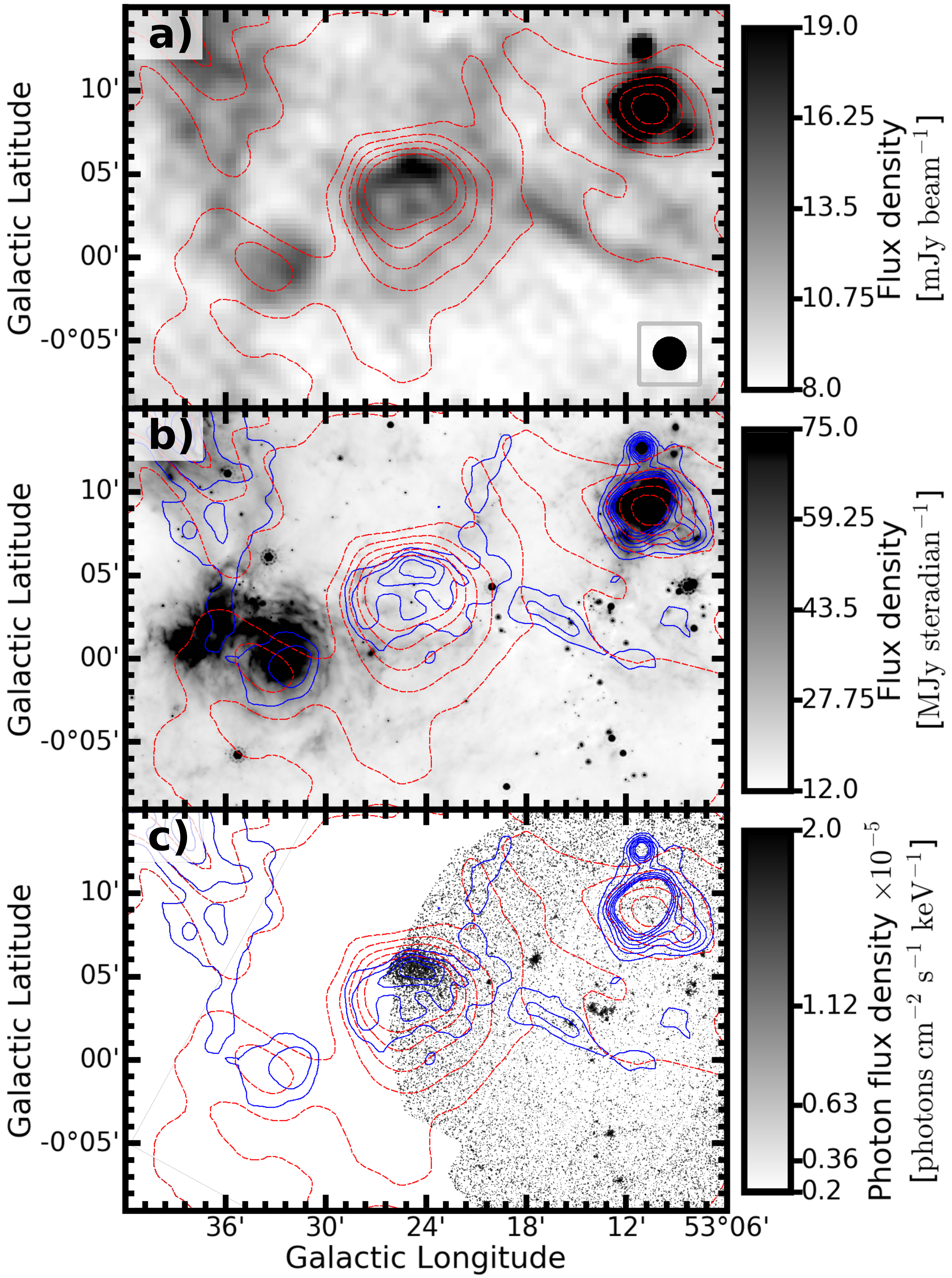}
\caption{Observations of $\mathrm{G}53.41+0.03$ at (a) $1.4\,\mathrm{GHz}$ using the VLA, (b) $24.0\,\mathrm{\mu m}$ using {\it Spitzer}, and (c) X-rays using \textit{XMM-Newton}. The dashed red contours are LOFAR HBA $144\,\mathrm{MHz}$ contours (contour levels: 0, 250, 500, 750, 1000 $\mathrm{mJy\,\,beam^{-1}}$) and the solid blue contours are VLA $1.4\,\mathrm{GHz}$ contours (contour levels: 12, 14, 16, 18, 20, 22, 24 $\mathrm{mJy\,\,beam^{-1}}$) from the image in (a). The VLA synthesised beam is shown in the bottom right corner of (a).
}
\label{fig: F5}
\end{figure*}

{In the LOFAR HBA and VLA observations $\mathrm{G}53.41+0.03$ has a shell- or bubble-like morphology which is brighter on the upper edge, as shown in Figure\,\ref{fig: F5}.} The radius of the shell at $144\,\mathrm{MHz}$ is $\sim5'$. As shown in Table\,\ref{tab: T1} $\mathrm{G}53.41+0.03$ has a radio spectral index of $\alpha=-0.6\pm0.2$. 

\subsection{X-ray results}

As described in Sec.\,\ref{xray_obs}, $\mathrm{G}53.84-0.75$ was observed by {\it ROSAT}. {We used the PIMMS\footnote{\url{https://heasarc.gsfc.nasa.gov/cgi-bin/Tools/w3pimms/w3pimms.pl}} tool with the optically thin plasma model APEC with temperature $0.3\,\mathrm{keV}$ and local Galactic absorption value of $2.4\times 10^{22}$~cm$^{-2}$ to obtain the $2\sigma$ upper limit for the flux. No X-ray feature coincident to the radio observations was detected. The $2\sigma$ upper limit for the absorbed/unabsorbed flux is F$_{0.4-2.4} \approx$ $2.4\times 10^{-13}$ / $4.1\times 10^{-11}$ erg~s$^{-1}$~cm$^{-2}$}.

\begin{figure}
\plotone{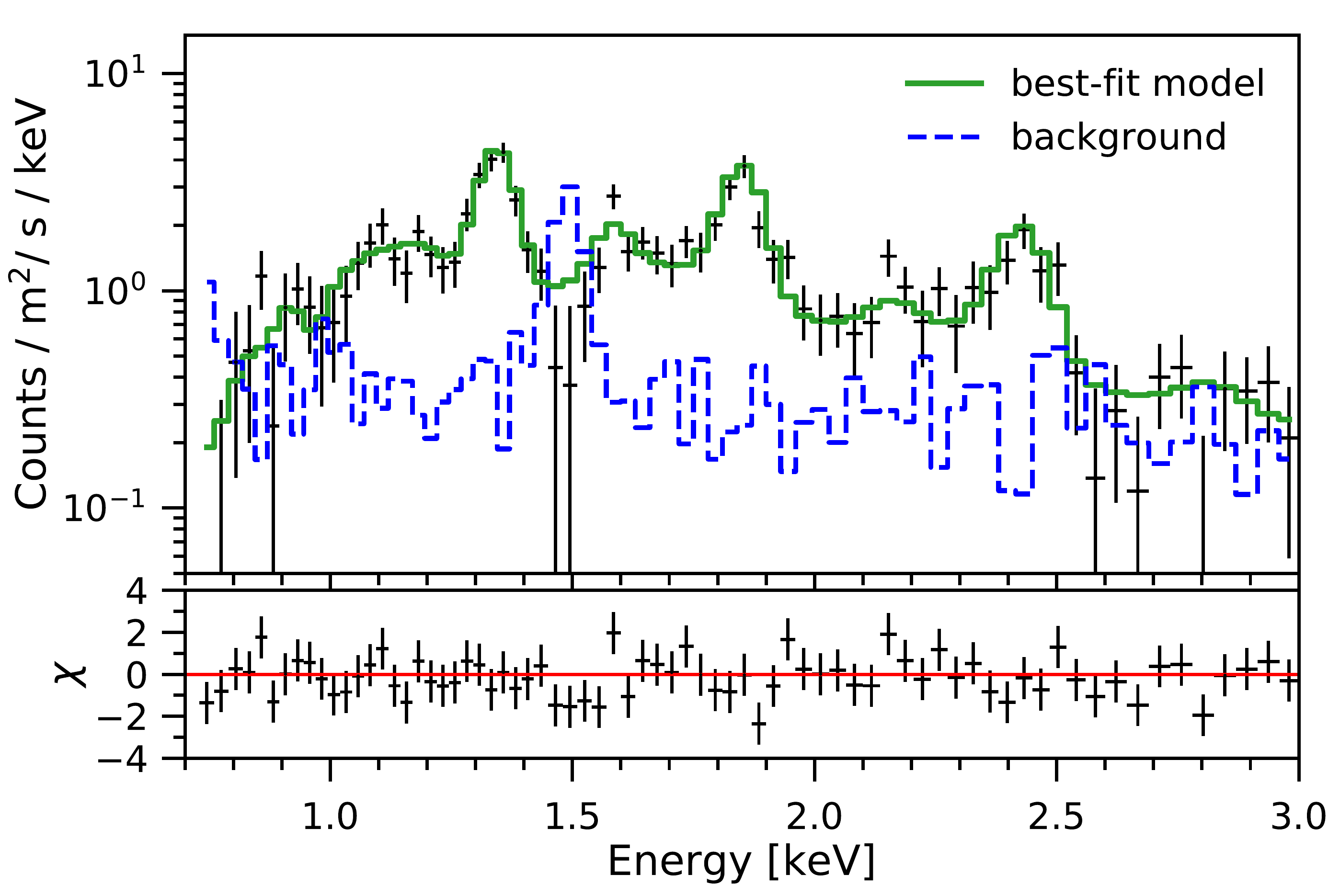}
\caption{X-ray spectrum with the best-fit model and residuals.\label{fig: F6}}
\end{figure}

\begin{figure}
\plotone{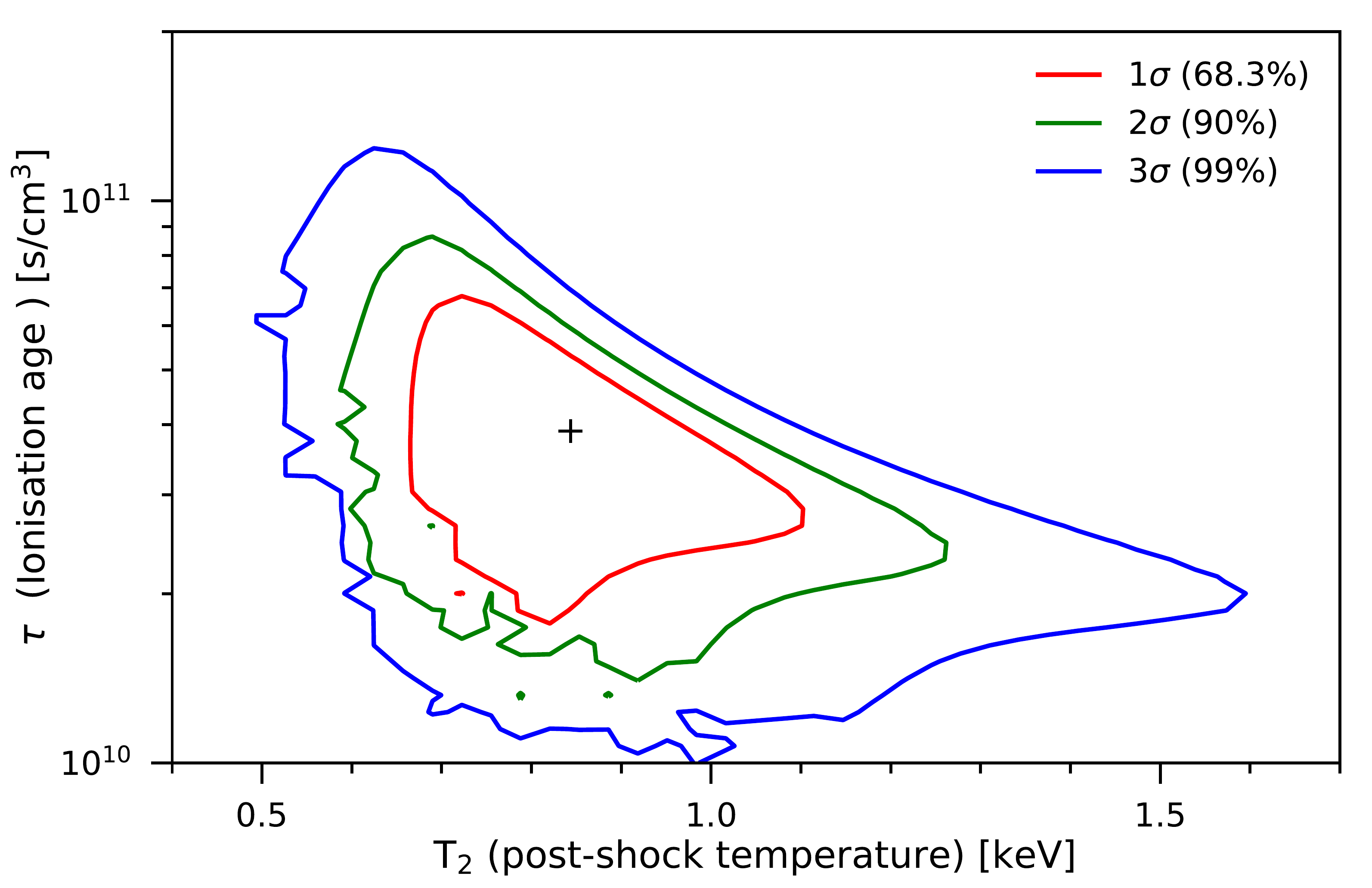}
\caption{Contour plot of the ionization age and post-shock temperature. Best fit values are indicated by a small cross. \label{fig: F7}}
\end{figure}

{The {\it ROSAT} and {\it XMM-Newton} X-ray observations of $\mathrm{G}53.41+0.03$} confirm the existence of an extended X-ray source at the location of $\mathrm{G}53.41+0.03$, particularly at the position of the radio-bright part of the shell\footnote{Since the spectral resolution of the {\it ROSAT} PSPC is poor and the images are noisy, we use only the \textit{XMM-Newton} observation for further analysis.}. The {\it XMM-Newton} X-ray spectrum  (Fig.~\ref{fig: F6}) shows  bright K-shell emission lines  from  magnesium, silicon, and sulfur and potential contributions from neon and iron around $1\,\mathrm{keV}$. This is typical of thermal emission from an optically thin plasma. The absorbed/unabsorbed flux of the source measured using {\it XMM-Newton} in the $0.7$ -- $3.0\,\mathrm{keV}$ energy range is F$=7.3 \times 10^{-13}$ / $3.1 \times 10^{-11}$ erg~s$^{-1}$~cm$^{-2}$. The best-fit NEI model is represented by a C-stat~/~d.o.f. of $83.48/64$. The parameters and $1\sigma$ errors are listed in Table \ref{tab: T2}, while the best fit model is shown in Figure \ref{fig: F6}.  The ionization age informs us how far out of ionization equilibrium the plasma  is, but given the narrow spectral range the parameter may correlate with the best-fit electron temperature $T_2$.  To test the robustness of our best fit ionization age we calculate the error ellipse of  $\tau$  and $T_2$, as shown in Figure \ref{fig: F7}.

\begin{table}
\centering
\begin{tabular}{lll|ll}
 Parameter & Unit  & Value &  Element & Abundance \\
\hline
\hline
\vspace{-5mm}
\\
N$_H$ & $10^{22}$ cm$^{-2}$ & 2.4$^{+0.2} _{-0.2}$ & Ne & 0.2$^{+0.7} _{-0.2}$ \\
n$_\mathrm{e} $n$_H V$ & 10$^{57}$ cm$^{-3}$  & 5$^{+2} _{-2}$ & Mg & 0.9$^{+0.3} _{-0.2}$  \\
T$_2$  & keV & 0.8$^{+0.2} _{-0.1}$  & Si & 0.5$^{+0.1} _{-0.1}$ \\
$\tau$ & $10^{10}$ s cm$^{-2}$ & 4$^{+2} _{-1}$ & S & 0.9$^{+0.2} _{-0.2}$ \\
& & & Fe & 1.3$^{+0.7} _{-0.5}$    \\
\hline
\hline
\multicolumn{4}{r}{Cstat/d.o.f} &
\multicolumn{1}{l}{83.48/64} \\
\end{tabular}
\caption{\textit{XMM-Newton} best-fit model results. The abundances are provided in Solar units.}
\label{tab: T2}
\end{table}

\subsection{Radio pulsation search results}

After performing a pulsation search as described in Sec.\,\ref{sec: obs pulse search} we found no
convincing astronomical signals in the data toward $\mathrm{G}53.41+0.03$, and we
ascribe the statistically significant signals that we did detect to RFI.

Given the non-detection of radio pulsations toward $\mathrm{G}53.41+0.03$, we can place
an upper limit on the integrated flux density of any associated radio pulsar.  We use
the modified radiometer equation \citep{1985ApJ...294L..25D}, and assume that interstellar
scattering does not have a significant effect on broadening the pulses
through multi-path propagation.  While the central ALFA beam has a
gain of $G \sim 10$\,K\,Jy$^{-1}$, the 6 outer beams have $G \sim
8$\,K\,Jy$^{-1}$.  We targeted the center of $\mathrm{G}53.41+0.03$ (specifically, RA$_{\rm J2000}$ = $19^h29^m57.41^s$, Dec$_{\rm J2000}$ = $+18^{\circ}09^{\prime}53.5^{\prime\prime}$) in a $T = 2400$-s
pointing with the central ALFA beam, which covered a region of roughly
1.6$^{\prime}$ in radius.  Since $\mathrm{G}53.41+0.03$ is roughly 10$^{\prime}$ wide, we also gridded a much larger
$\sim 10^{\prime}$ wide region around $\mathrm{G}53.41+0.03$ in case the pulsar has
moved from its birth site near the center of the SNR.  In our sensitivity calculations we thus
consider two scenarios: 1) where the pulsar is close to the center of
$\mathrm{G}53.41+0.03$, and where we should use $G = 10$\,K\,Jy$^{-1}$ and $T = 2400$\,s
and 2) a scenario in which the pulsar is offset by several arcminutes,
and where $G = 8$\,K\,Jy$^{-1}$ and $T = 900$\,s.  Furthermore, if
the pulsar is located towards the half-power
sensitivity point of one of the beams, then the effective sensitivity
is also half.  We make this conservative assumption for scenario 2.

The receiver temperature $T_{\rm rec}$ = 25\,K and the sky temperature
in this direction of the Galactic plane is $T_{\rm sky}$ = 5\,K at 1400\,MHz.  We
assume a W = 10\% pulse duty cycle and a signal-to-noise S/N = 10 for
detection.  The two orthogonal linear polarizations, $n_{\rm p}$, of
the receiver were summed, and the appropriate bandwidth is $\Delta\nu
= 172$\,MHz.  Finally, using the modified radiometer equation, and
assuming no additional losses due to digitization, we find for
scenario 1:

\begin{equation}
S^1_{\rm max} {\rm [mJy]} \frac{S/N (T_{\rm rec} + T_{\rm sky})}{G \sqrt{T \Delta\nu n_{\rm p}}} \sqrt{\frac{W}{(1-W)}} = 0.011\,{\rm mJy}
\end{equation}

For scenario 2, where the putative pulsar is more offset from $\mathrm{G}53.41+0.03$,
$S^2_{\rm max} = 0.045$\,mJy.  These are deep upper-limits on the flux
density of any pulsar associated with $\mathrm{G}53.41+0.03$.  Of the known young
pulsars in the ATNF catalog, only a few have lower measured radio flux density \citep{2005AJ....129.1993M}.  However, because of beaming and the possibility of significant interstellar
scattering, these limits do not definitively exclude a young pulsar associated with
$\mathrm{G}53.41+0.03$.

\section{Discussion}
\label{sec: discussion}

{Here we will discuss the characteristics and nature of each SNR candidate. We will focus on $\mathrm{G}53.41+0.03$, including calculating its approximate distance and age.}

{\textbf{G51.21$+$0.11:} SNR candidate $\mathrm{G}51.21+0.11$ has a negative spectral index, $\alpha=-0.7\pm0.21$, and a complex morphology coincident with a known HII region. There are no \textit{XMM-Newton} or \textit{Chandra} observations in the direction of the candidate to confirm its nature. We find $\mathrm{G}51.21+0.11$ to be an interesting object that is possibly an SNR, but further investigation using X-ray observations is required.}

{\textbf{G52.37$-$0.70:} Although $\mathrm{G}52.37-0.70$ has a shell-like morphology in the VLA observations, it has a spectral index of $\alpha=0.3\pm0.3$ fitted using the VLA and WSRT integrated flux densities. The spectral index indicates that this candidate is unlikely to be an SNR, and as such the {\it Fermi} source within the radius of the candidate (see Sec.\,\ref{sec: high energy obs}) is unlikely to be associated.}

{\textbf{G53.07$+$0.49:} Candidate $\mathrm{G}53.07+0.49$ has a small angular size in the VLA observations, but the peak flux density in the WSRT and LOFAR observations is offset from the SNR candidate location suggested by \citep{2017arXiv170510927A}. As such we do not measure WSRT or LOFAR flux densities for this candidate, and as there are no X-ray observations available, further investigation using X-ray or higher resolution low-frequency observations is required to comment on the nature of this candidate.}

{\textbf{G53.84$-$0.75:} It is not clear what emission is SNR candidate $\mathrm{G}53.84-0.75$ and there are large errors on the VLA integrated flux density from \citet{2017arXiv170510927A}. This, as well as the strange spectral shape, suggests that there is no discrete, extended object at this position. This is supported by the {\it ROSAT} X-ray non-detection. For this reason we find it unlikely that $\mathrm{G}53.84-0.75$ is an SNR.}

\textbf{G54.1$+$0.3:} Whether PWN $\mathrm{G}54.1+0.3$ has an SNR shell has been in question since \citet{2010ApJ...709.1125L} found faint radio emission around the PWN{, which is just visible in the VLA observation (Fig.\,\ref{fig: F4}).} \citet{2002ApJ...568L..49L} found no evidence of a shell in their {\it Chandra} observations, while \citet{2010A&A...520A..71B} found hints of a very faint, diffuse shell using {\it Suzaku} and {\it XMM-Newton}. \citet{2017arXiv170510927A} find that the shell suggested by \citet{2010ApJ...709.1125L} is more likely to be part of the surrounding HII region. Alternatively, \citet{2017arXiv170510927A} suggest a slightly smaller radius shell ($7.2'$) as a possible shell around PWN $\mathrm{G}54.1+0.3$ with an integrated flux density of $1.46\,\mathrm{Jy}$ at $1.4\,\mathrm{GHz}$. There is no evidence for extended emission around PWN $\mathrm{G}54.1+0.3$ in our LOFAR HBA observation, as can be seen in Figure \ref{fig: F4} (bottom panel), aside from the known HII region $\mathrm{G}053.935+0.228$ \citep{2014ApJS..212....1A}. This is supported by the low flux-densities measured by WSRT and LOFAR (shown in Table\,\ref{tab: T1}) using a region of radius $7.2'$ and subtracting the flux density of the PWN. We find it unlikely that there is a shell around PWN $\mathrm{G}54.1+0.3$.

{\textbf{G53.41$+$0.03:} $\mathrm{G}53.41+0.03$ has a morphology common to SNRs.} Using the flux densities shown in Table\,\ref{tab: T1} we find that the $\mathrm{G}53.41+0.03$ has a steep negative radio spectral index, $\alpha=-0.6\pm0.2$, as expected for an SNR. X-ray analysis indicates that the plasma of $\mathrm{G}53.41+0.03$ has a relatively high temperature of $\mathrm{T}_2\sim 0.8\,\mathrm{keV}$. The ionization age $\tau\sim10^{10.6}\,\mathrm{s}\cdot\mathrm{cm}^{-3}$ is much lower than needed for ionization/recombination balance ($\tau\geq10^{12}\,\mathrm{s}\cdot\mathrm{cm}^{-3}$). The fact that the spectrum is far out of ionization equilibrium is a clear signature that the source is an SNR \citep{Vink2012}, as no other known source class has gas tenuous enough and/or is young enough to be far out of ionization equilibrium. {We therefore confirm that $\mathrm{G}53.41+0.03$ is an SNR, and further investigate it by calculating its approximate distance and age.}

\subsection{The distance to G53.41+0.03}
\label{sec: G53 distance}
Estimating the distance to Galactic SNRs is notoriously difficult. There are few methods  that give reliable results, such as kinematic methods, based on  optical Doppler shifts combined with  proper motion of optical filaments \citep[e.g.][for Cas A]{1995ApJ...440..706R}, or, less reliably, 
21cm line absorption combined with a Galactic  rotation model \citep[see e.g.][for an explanation and SNR application of the model]{2009ApJ...699.1153R,2012ApJ...746L...4K}.
In contrast, SNRs located in the Magellanic Clouds can be reliably placed
at the distance of these satellite galaxies.  By using reliable distance estimates some secondary distance indicators have been developed, such as
the X-ray Galactic
absorption column \citep{1994MNRAS.268L...5S} and the $\mathrm{\Sigma-D}$ relation \citep{2014SerAJ.189...25P}.

A first indication of the distance of an SNR can be its positional association with  a spiral arm.  However, the reason that the investigated field  is so rich in sources is that the line of sight crosses the Sagittarius-Carina arm tangentially as well as regions of the Perseus arm. Taking the Galactic spiral arm model of \citet{2009A&A...499..473H}, we  find that the  $l=53.4^{\circ}$ line of sight intercepts the Sagittarius arm \citep[arm -3 in][]{2009A&A...499..473H} between $\sim 4$~kpc and 7.5~kpc, and the Perseus arm at 9.6~kpc. Given that the Sagittarius arm is  tangential along the line of sight, this suggests a probable distance between 4.5 and 7.5 kpc.

\citet{1994A&A...288L...1S} derived a relation between column density and distance of $N_\mathrm{H}=8.4\times 10^{21} d^{1.58}$~cm$^{-2}$. The measured
 column density of  $N_\mathrm{H}=2.4\times 10^{22}$~cm$^{-2}$ (Table~\ref{tab: T2}), therefore, suggests a distance of $\sim 8.4$~kpc.
However, one should be cautious here, because the line of sight crosses the arm tangentially, which  is likely to lead to a column density that is higher than average  for a given distance.

{The surface brightness of $\mathrm{G}53.41+0.03$ normalized to 1 GHz is  $\Sigma = 8.3\times 10^{-21}$~W\ m$^{-2}$ Hz$^{-1}$sr$^{-1}$. The $1\,\mathrm{GHz}$ surface brightness was obtained using the $1.4\,\mathrm{GHz}$ flux density measured by \citet{2017arXiv170510927A} and a spectral index of $\alpha=-0.6$ (see Tab.\,\ref{tab: T1}). Using the relation between diameter and surface brightness (the $\mathrm{\Sigma-D}$ relation) in \citet{2014SerAJ.189...25P} gives yet another distance estimate of $8\,\mathrm{kpc}$. However, we know that the $\mathrm{\Sigma-D}$ relation is controversial, as there is large scatter which may relate to the SNR environments, and there is debate on the statistical validity of the relation \citep[e.g.][]{2005MNRAS.360...76A,2005MNRAS.364..217F,2014BASI...42...47G}.}

{The distance estimates based on the X-ray absorption and $\mathrm{\Sigma-D}$ relation, although uncertain, are consistent with the idea that the SNR is located in the Sagittarius-Carina arm, but suggest that the SNR is on the far-side of the arm.} We therefore adopt a distance of $7.5\,\mathrm{kpc}$ for $\mathrm{G}53.41+0.03$.
The angular radius of $\sim 5^\prime$ translates then
into a physical radius of 10.7$d_{7.5}$~pc,
with $d_{7.5}$ the distance in units of $7.5\,\mathrm{kpc}$.

\subsection{The age of G53.41+0.03}
\label{sec: G53 age}

{The spectrum of $\mathrm{G}53.41+0.03$ allows us to put some constraints on the density and age of the SNR. To do this we need a volume estimate.} Given a typical volume filling fraction of 25\%\footnote{A strong shock has a compression factor of 4. This means that roughly 25\% of the volume, approximated by a sphere, will emit.} and assuming a spherical morphology, we estimate the volume to be $V_\mathrm{SNR}=3.3\times 10^{58}d_{7.5}^3$~cm$^{3}$. The X-ray spectrum was obtained for only $\sim20$\% of the shell, so we take $V_\mathrm{X}\approx 6.7\times 10^{57}d_{7.5}^3$~cm$^{3}$ to be the volume pertaining to the X-ray spectrum. Taking
$n_\mathrm{e}\approx 1.2n_{H}$ in the emission measure $n_\mathrm{e} n_H V$, we obtain the density $n_\mathrm{H}\approx 0.8 d_{7.5}^{-3/2}$~cm$^{-3}$. 
Using this number together with the best-fit ionization age of $n_\mathrm{e}t=4\times 10^{10}$~cm$^{-3}$s we find an approximate age
of 1600$d_{7.5}^{3/2}$~yr.

The measured electron temperature corresponds to a shock  velocity of $V_\mathrm{s}\approx 800$~km\,s$^{-1}$ or higher if the electron temperature is lower than the ion temperature \citep{Vink2012}.  For the Sedov-Taylor self-similar evolution model we have $V_\mathrm{s}=0.4 R/t$.Using $R=10.7 d_{7.5}$~pc,  gives then an approximate age  of $\sim 5300d_{7.5}$~yr. Using the Sedov-Taylor  evolution model of $R^5=2.026 Et^2/\rho$, with $E=10^{51}$~erg gives yet another estimate of the age of  $\sim 7800d_{7.5}^{7/4}$~yr.
The two estimates based on the Sedov-Taylor model give roughly similar results for the canonical
explosion energy of $10^{51}$~erg ($t\approx 6500 \pm 1500$~yr), whereas the estimate
based on the ionization age suggests a younger age. This discrepancy may be due to non-standard
evolution scenarios, for example evolution in a wind-blow cavity. This needs to be
addressed in follow-up studies. However, these estimates agree that $\mathrm{G}53.41+0.03$
is an SNR with an age somewhere between 1000 and 8000~yr.
X-ray observations centered on and covering the whole SNR are needed to fully characterize the properties of $\mathrm{G}53.41+0.03$.

\section{Conclusion}
\label{sec: conclusion}

We confirm that SNR candidate, $\mathrm{G}53.41+0.03$, is in fact an SNR using {\it XMM-Newton} observations, and LOFAR observations targeting PWN $\mathrm{G}54.1+0.3$. $\mathrm{G}53.41+0.03$ has a shell-like morphology in the radio, with a radius of $\sim 5\,'$. Using LOFAR HBA observations, as well as archival WSRT and VGPS mosaics, we confirm that $\mathrm{G}53.41+0.03$ has a steep spectral index ($\alpha=-0.6\pm0.2$), typical of synchrotron radiation from SNRs. MIPSGAL observations show that $\mathrm{G}53.41+0.03$ has no IR component. Archival \textit{XMM-Newton} observations show that $\mathrm{G}53.41+0.03$ has an associated X-ray component with a coincident morphology to the radio shell. Furthermore, analysis and fitting of the \textit{XMM-Newton} observation show that $\mathrm{G}53.41+0.03$ has strong emission lines and is best characterized by a non-equilibrium ionization model, with an ionization age and normalization typical for an SNR with an age between 1000 and 8000 yr and a density of $n_\mathrm{H}\approx 0.8 d_{7.5}^{-3/2}$~cm$^{-3}$. Given the X-ray, IR, and radio characteristics of $\mathrm{G}53.41+0.03$, we confirm that it is a new Galactic Plane SNR. We do not find a pulsar associated with $\mathrm{G}53.41+0.03$, but the upper-limits on the flux density do not exclude the possibility of a young pulsar that is exceptionally weak or not beamed towards Earth.

We also investigate five other SNR candidates from \citet{2017arXiv170510927A} in the same LOFAR FoV. We show that three of these candidates ($\mathrm{G}52.37-0.70$, $\mathrm{G}53.84-0.75$ {and the shell around PWN $\mathrm{G}54.1+0.3$}) are unlikely to be SNRs and one, $\mathrm{G}51.21+0.11$, is a good SNR candidate that requires further investigation. This demonstrates that it is important to further investigate SNR candidates using low-frequency observations with telescopes such as WSRT and LOFAR.

\acknowledgments

We would like to thank Vincent Morello, George Heald, Raymond Oonk, Andre Offringa, Jess Broderick, Pedro Salas, Alex Mechev, and Irene Polderman for useful discussions and assistance with LOFAR imaging and calibration.
LND and JWTH acknowledge support from the European Research Council under the European Union's Seventh Network Framework Programme (FP/2007-2013) / ERC Grant Agreement nr. 337062. JWTH is an NWO Vidi fellow.
This paper is based (in part) on data obtained with the International LOFAR Telescope (ILT) under project code LC4\_011. LOFAR \citep{2013A&A...556A...2V} is the Low Frequency Array designed and constructed by ASTRON. It has observing, data processing, and data storage facilities in several countries, that are owned by various parties (each with their own funding sources), and that are collectively operated by the ILT foundation under a joint scientific policy. The ILT resources have benefited from the following recent major funding sources: CNRS-INSU, Observatoire de Paris and Universit\'e d'Orl\'eans, France; BMBF, MIWF-NRW, MPG, Germany; Science Foundation Ireland (SFI), Department of Business, Enterprise and Innovation (DBEI), Ireland; NWO, The Netherlands; The Science and Technology Facilities Council, UK.
This paper is also based (in part) on observations obtained with {\it XMM-Newton}, an ESA science mission with instruments and contributions directly funded by ESA Member States and NASA.

\vspace{5mm}
\facilities{LOFAR, VLA, WSRT, \textit{XMM-Newton}, \textit{ROSAT}, {\it Spitzer}}

\software{Numpy \citep{2011arXiv1102.1523V}, Astropy \citep{2013A&A...558A..33A}, Aplpy \citep{2012ascl.soft08017R}, WSClean \citep{offringa-wsclean-2014}, pyBDSF\footnote{\url{http://www.astron.nl/citt/pybdsm/index.html}}, DS9\footnote{\url{http://ds9.si.edu/site/Home.html}}, SAS v14.0\footnote{\url{https://www.cosmos.esa.int/web/xmm-newton}}, Spex v3.04 \citep{Kaastra1996}, DPPP}

\end{document}